\documentclass{aip-cp}

\pdfoutput=1

\usepackage[numbers]{natbib}
\usepackage{rotating}
\usepackage{graphicx}
\usepackage{aas_macros}
\begin{document}

\title{VERITAS and VLBA Observations of HESS J1943+213}

\author[aff1]{Karlen Shahinyan}
\eaddress{shahin@astro.umn.edu}
\author{the VERITAS Collaboration}
\eaddress[URL:]{http://veritas.sao.arizona.edu}

\affil[aff1]{Minnesota Institute for Astrophysics, School of Physics and Astronomy, University of Minnesota, MN, 55455, USA}

\maketitle

\begin{abstract}
HESS J1943+213 is a very high energy (VHE; $>$100 GeV) $\gamma$-ray source in the direction of the Galactic plane. 38 hours of deep VERITAS observations taken over two seasons detect the source with $\sim$20 $\sigma$ significance. Monitoring observations of HESS J1943+213 show a remarkably stable flux and spectrum in VHE $\gamma$-rays. Studies exploring the classification of HESS J1943+213 are converging towards accepting the source as an extreme synchrotron BL Lac object. Specifically, overall SED characteristics of the source, the detection of a potential host galaxy in near-IR imaging, and VLBI observations of the HESS J1943+213 radio counterpart showing extended jet-like emission at milliarcsecond scale and core flux density variability establish the source as a blazar. Recent Very Long Baseline Array (VLBA) observations of the source (shown here for the first time) confirm the extended structure found in the 1.6 GHz band and detect the jet-like component in the 4.6 GHz and the 7.3 GHz bands. The spectral indices of the core and the jet-like components derived from the VLBA observations are in a range typical for blazars. HESS J1943+213 fits the extreme HBL description; however, with variability detected only at milliarcseconds scales in radio, it appears to be an abnormally stable VHE $\gamma$-ray blazar.
\end{abstract}

\section{Background}

Blazars are the most common extragalactic objects in the very-high-energy (VHE; $>$ 100~GeV) $\gamma$-ray sky with the current count at 62\footnote{http://tevcat2.uchicago.edu}. Their emission is characterized by a spectral energy distribution (SED) with a double-humped structure. Under the synchrotron-self-Compton (SSC) description the lower energy hump is produced by synchrotron emission from a population of leptons, while the higher energy hump is attributed to inverse-Compton emission from the interactions of these synchrotron photons with the same population of leptons~\cite{konigl}.
Based on the location of the peak of the synchrotron emission, blazars are classified into low, intermediate, and high synchrotron peak BL Lac objects (LBL, IBL, and HBL respectively). With no apparent cutoff in the synchrotron emission up to 195 keV, HESS J1943+213 belongs to a subclass of HBLs with peak locations at \textgreater1 keV energies known as extreme HBLs~\cite{cost}. VHE-detected extreme HBLs are rare with only a handful of examples, including 1ES 0229+200, 1ES 0502+675, and 1ES 0347-121.

Because of their X-ray and $\gamma$-ray spectra, extreme HBLs are thought to have the highest bulk Lorentz factors and Doppler boosting factors compared to other blazars in the one-zone SSC paradigm. However, VLBI measurements in the radio have shown low bulk Lorentz and Doppler factors in these objects, leading to proposals of various multi-component jet models or multiple emission zones (\cite{PE} and references therein). Extreme blazars are also intriguing within the context of blazar unification efforts ~\cite[e.g.,][]{meyer}, where extreme HBLs are expected to have the weakest jets and the smallest viewing angles. 

Due to the hard VHE emission, the sharpness of the $\gamma$-ray peak, and a stable flux extreme HBLs are prime tools for setting lower limits on the strength of intergalactic magnetic fields (IGMF) \cite[e.g.,][]{tevecchio,tanaka14,arlen}. Moreover, the VHE spectra of extreme HBLs can be used to study $\gamma$-ray propagation effects from measuring the density of the extragalactic background light to exploring exotic mechanisms such as photon oscillations into axion-like particles, Lorentz invariance violation, and production of secondary $\gamma$-rays from electromagnetic cascades initiated by ultra-high-energy cosmic-rays in extragalactic space. Extreme BL Lacs are especially appealing candidates for the latter scenario because of the lack of detected variability over timescales shorter than a month~\cite{TB}.

HESS J1943+213 is a recent addition to this blazar class after the debate over whether it is a blazar or a Galactic source was settled in favor of the former. 
Despite the favored blazar classification from the time of the discovery \cite{hess11}, there were initial challenges primarily from 1.6 GHz VLBI observations with EVN measuring a brightness temperature too low for an AGN \cite{gabanyi13}. The blazar classification was solidified by the combination of the discovery of a potential host galaxy in the near-IR \cite{peter14}, a comparative study of the source SED with another extreme HBL \cite{tanaka14} and follow up VLBI observations that measured a  brightness temperature within the blazar range, a jet-core structure, and overall radio properties consistent with a blazar classification~\cite{akiyama16}. For a blazar, HESS J1943+213 is exceptionally stable. The only claims of flux variability for the source come from comparisons of VLBI-measured flux densities \cite{akiyama16,straal16}; whereas, no significant variability has yet been detected in multi-year X-ray and $\gamma$-ray observations~\cite{hess11,shahinyan15}.

The precise distance to HESS J1943+213 is not known, though indirect limits from imaging of its potential host galaxy and its $\gamma$-ray spectra place it in the 0.03 $<$ z $<$ 0.45 redshift range \cite{peter14}. 

\section{Observations, Results, and Discussion}

\subsection{VERITAS}

The Very Energetic Radiation Telescope Array System (VERITAS) is an imaging atmospheric Cherenkov telescope (IACT) array located at the Fred Lawrence Whipple Observatory (FLWO) in southern Arizona (31$^{\circ}$40$'$N, 110$^{\circ}$57$'$W,  1.3~km a.s.l.). VERITAS is composed of four 12-m telescopes, each equipped with a 499 photo-multiplier tube camera providing a 3.5$^{\circ}$ field of view. The array can reliably reconstruct $\gamma$-rays with energies between 85~GeV and $>$30~TeV and has an angular resolution (at 68$\%$ containment) of $<$ 0.1 degrees for a 1~TeV photon. The energy resolution is 17$\%$ at 1~TeV, with a 10$^{5}$ m$^{2}$ peak effective area \cite{holder06}.

\begin{figure}[!h]
 \centerline{\includegraphics[width=300pt]{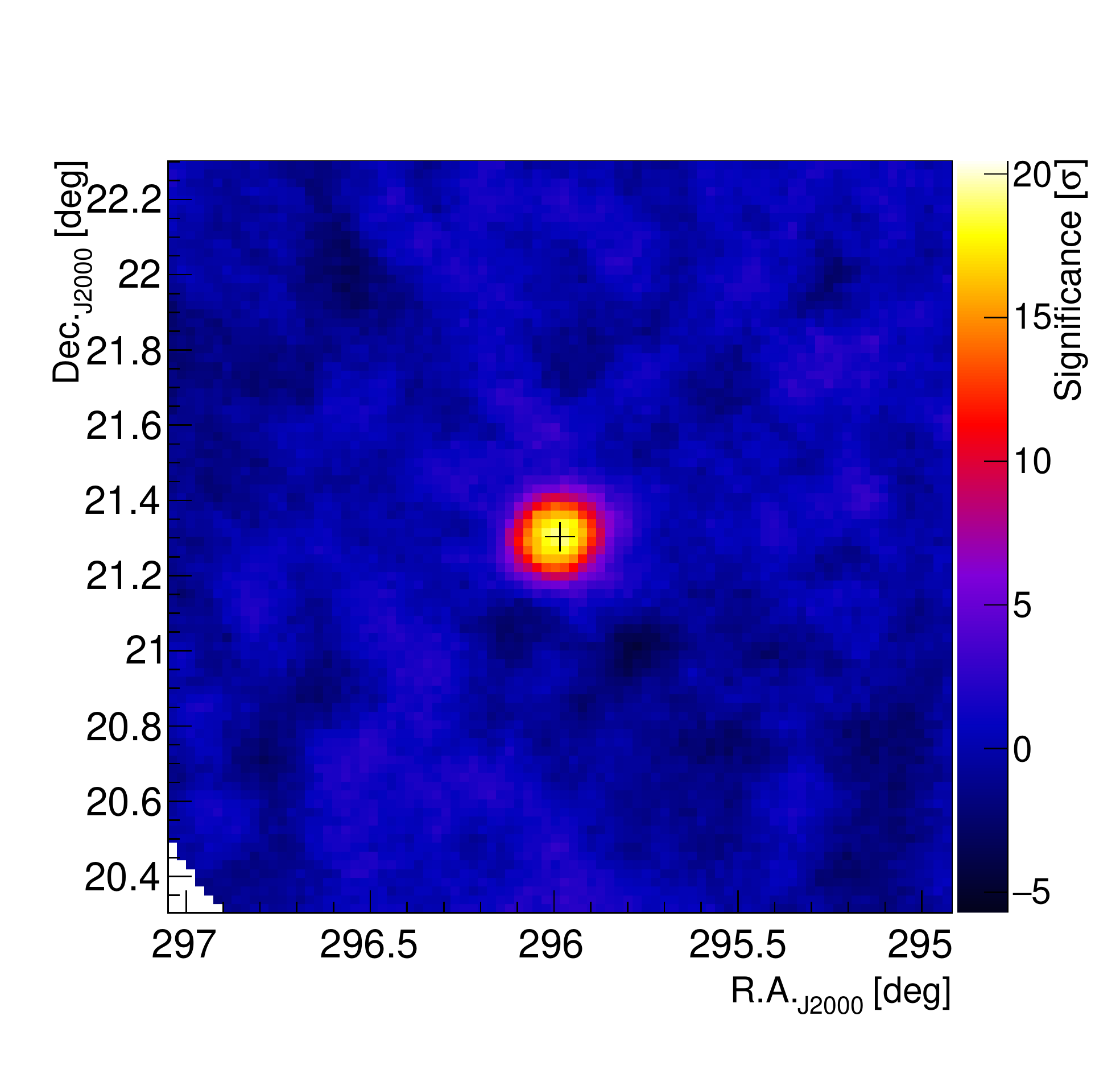}}
  \caption{VERITAS significance map, centered on the HESS J1943+213 position.\label{fig:veritas_skymap}}
\end{figure}

VERITAS has observed HESS J1943+213 for a total exposure time of 37.2~hours, amounting to a weather-cleaned live time of 30.9~hours, with initial results from the first year of observations reported in~\citet{shahinyan15}. The source elevation is within the 63$^{\circ}$ -- 80$^{\circ}$ range, with the common low energy threshold determined to be 178~GeV. 
Using the full dataset, the source is detected with an excess of 19.3 $\sigma$ centered at $\alpha$=19h43$'$59$''$ and $\delta$=21$^{\circ}$19$'$05$''$, consistent with the catalog position of HESS J1943+213. The VERITAS significance map from these observations, included in Figure~\ref{fig:veritas_skymap}, is clean showing no other sources in the field of view.

\begin{figure}[!h]
 \centerline{\includegraphics[width=300pt]{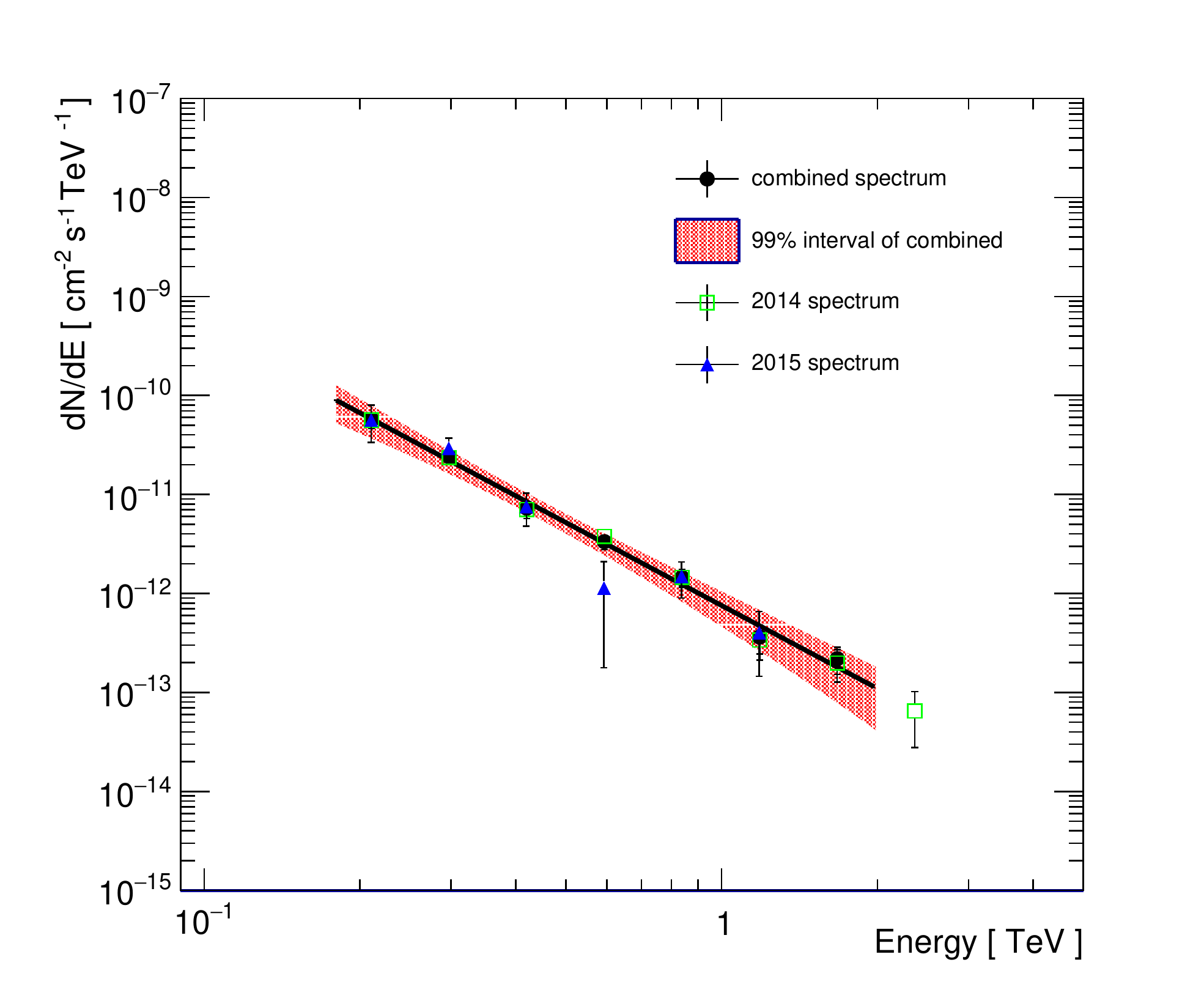}}
  \caption{VERITAS spectra of HESS J1943+213 from multiple epochs.\label{fig:veritas_spectra}}
\end{figure}

The spectra of HESS J1943+213 are constructed separately for each year of observations, as well as for the entire dataset and are presented in Figure~\ref{fig:veritas_spectra}. The spectra from individual years agree with each other within the statistical uncertainties, indicating no evidence for spectral
variability on a yearly timescale. The combined spectrum is fit well by a power-law function with a spectral index of -2.8$\pm$0.1$_{\mathrm{stat}}$ in the 
200~GeV--2~TeV energy range. This is consistent with the spectral index of the source reported by H.E.S.S., $\Gamma$ = -3.1$\pm$0.3$_{\mathrm{stat}}$$\pm$0.2$_{\mathrm{sys}}$ between 470 GeV and 6 TeV \cite{hess11} (in the same energy range, the VERITAS-measured spectral index is -2.85$\pm$0.3$_{\mathrm{stat}}$).

\begin{figure}[!h]
 \centerline{\includegraphics[width=400pt]{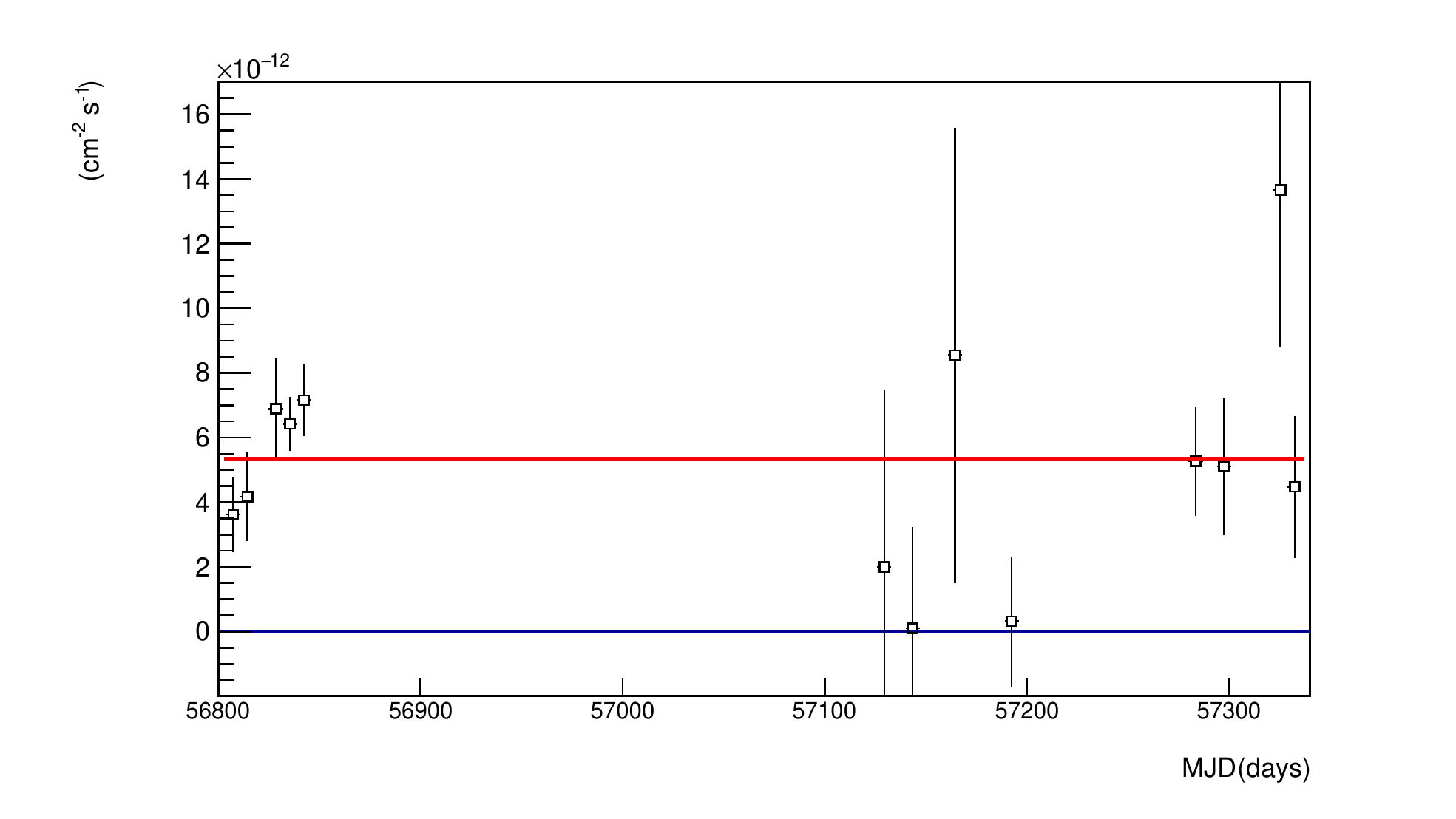}}
  \caption{VERITAS light curve of HESS J1943+213.\label{fig:veritas_lightcurve}}
\end{figure}

Figure~\ref{fig:veritas_lightcurve} illustrates the weekly-binned VERITAS light curve of HESS J1943+213. The
average flux of the source is (5.35$\pm$0.44)$\times$10$^{-12}$ cm$^{-2}$ s$^{-1}$ above 200 GeV. A constant line fit to the light curve gives a reduced $\chi^2$ = 1.74, corresponding to a p-value of 0.05 for a constant flux.
H.E.S.S. reported a source flux of (1.3$\pm$0.2$_{\mathrm{stat}}$$\pm$0.3$_{\mathrm{sys}}$)$\times$10$^{-12}$  cm$^{-2}$ s$^{-1}$
 above 470~GeV \cite{hess11} from observations taken between 2005 and 2008. This is consistent with the VERITAS flux above 470~GeV of (1.47$\pm$0.16)$\times$10$^{-12}$ cm$^{-2}$ s$^{-1}$. Consequently, in addition to the remarkable stability of the source flux within two years of VERITAS observations, there is
also good agreement between flux from observations more than six years apart from two independent VHE $\gamma$-ray instruments.

\subsection{Very Long Baseline Array}

The radio counterpart to HESS J1943+213 was  observed with the Very Long Baseline Array (VLBA) on 2015 August 11 (Project ID: BS246)
as part of a request to follow up on the European VLBA Network (EVN) detection and characterization of the source \cite{gabanyi13}. The observations were targeted at the position reported from the EVN detection: $\alpha$ = 19$^{h}$43$'$56.$''$2372, Dec:  21$^{\circ}$18$'$23.$''$402.
All 10 VLBA antennae participated in the observations: Mauna Kea (Hawaii), Owens Valley (California), Brewster (Washington), North Liberty (Iowa), Hancock (New Hampshire), Kitt Peak (Arizona), Pie Town (New Mexico), Fort Davis (Texas), Los Alamos (New Mexico), St. Croix (Virgin Islands). The total length of the observations was 4 hours, which included exposures on a phase calibrator source, J1946+2300 and a bandpass calibrator source, 3C 345. 

\begin{figure}[!h]
 \centerline{\includegraphics[width=500pt]{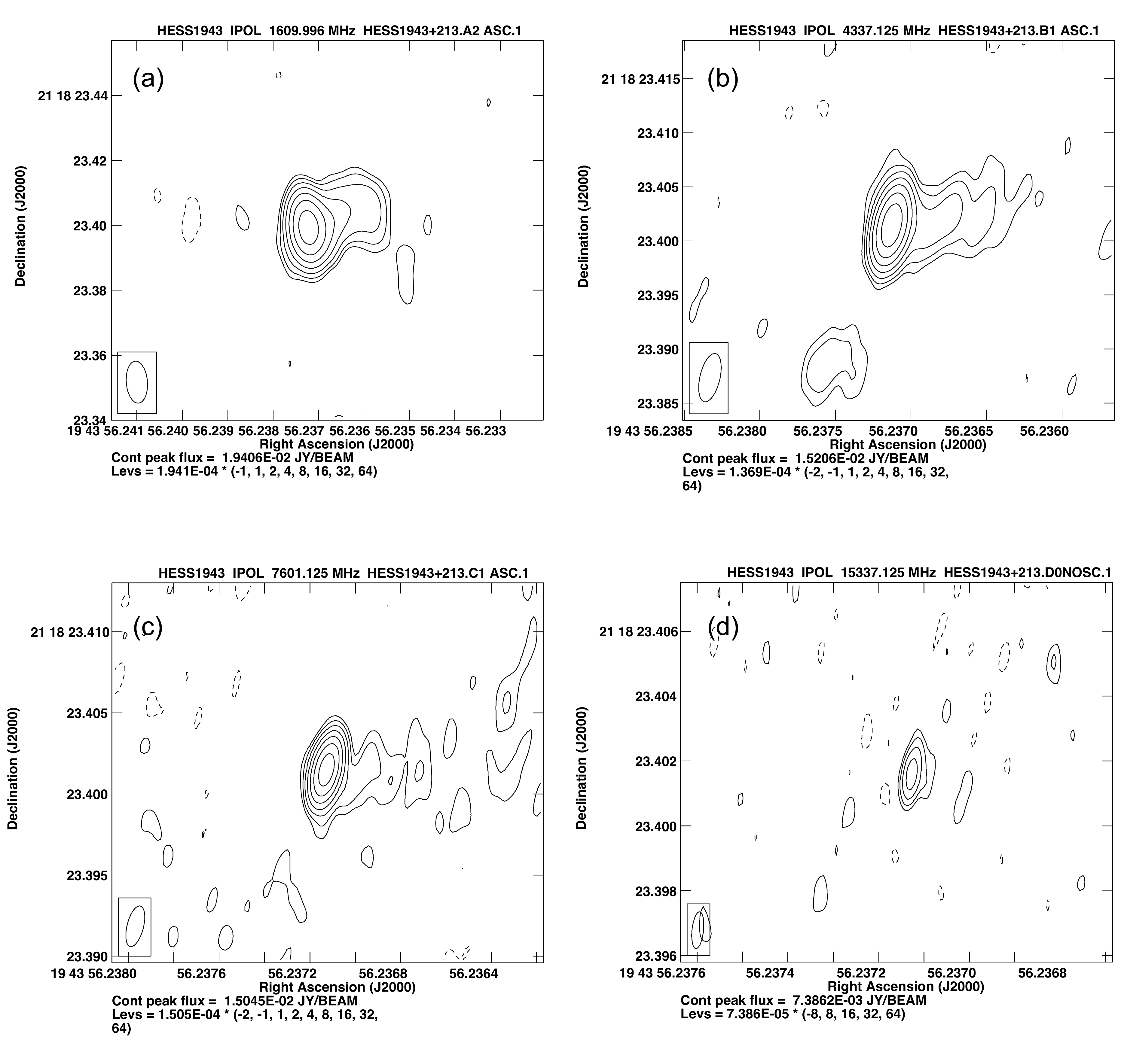}}
  \caption{Contour images of HESS J1943+213 in VLBA 1.6 GHz (a), 4.3 GHz (b), 7.6 GHz (c), and 15 GHz (d) bands.\label{VLBA_images}}
\end{figure}

The NRAO Astronomical Image Processing System (AIPS) is used to reduce and calibrate the VLBA data for HESS J1943+213. Images were produced for each band and are displayed in Figure~\ref{VLBA_images}. There is clear evidence for extended, jet-like emission in the 1.6 GHz, 4.3 GHz, and 7.6 GHz images. This is the first detection of the extended milliarcsecond-scale structure of this source in 4.3 GHz and 7.6 GHz bands, allowing multifrequency exploration of the source properties in VLBI. The core-jet structure has been previously detected through a deep 1.6~GHz band observations with EVN \cite{akiyama16}.

The core brightness temperature (T$_{B}$) of the HESS J1943+213 counterpart is estimated using images from all bands except for the 15 GHz band, where the sensitivity was too low for phase and amplitude self-calibration. Lower limits  are derived due to a partially resolved core and the possibility of interstellar scattering resulting in T$_{B}$ $>$~1.1$\times$10$^{9}$~K, 2.0$\times$10$^{9}$~K, and 1.6$\times$10$^{9}$~K for the 1.6 GHz, 4.3 GHz, and 7.6 GHz bands respectively. These values are well within the range for blazars.  We do not confirm the significantly lower brightness temperature measurement of T$_{B}$ = 7.7$\times$10$^{7}$~K, which was based on the EVN 1.6 GHz image \cite{gabanyi13} and which has subsequently been reanalyzed showing higher brightness temperature, T$_{B}$ $>$ 1.8$\times$10$^{9}$ \cite{akiyama16}. 

A spectral index of -0.3 $\pm$ 0.06 is measured for the HESS J1943+213 core, determined from all 4 bands, and an index of -1.1 $\pm$ 0.4 for the extended emission is calculated over 9.3 square milliarcseconds based on the 4.3 GHz and 7.6 GHz images. Both measurements are consistent with reported values for blazar cores and jets from the MOJAVE sample \cite{hovatta14}, but they are discrepant from the only available radio spectral index measurements of the core of this source from e-MERLIN that find an index of -0.03 $\pm$ 0.03 \cite{straal16}. The e-MERLIN observations do not resolve the source, however, and the core spectral index calculation using these observations is affected by emission from the jet-like feature.

Based on our position measurement of the HESS J1943+213 counterpart and the position reported from the \citet{gabanyi13} EVN detection, we see a change in position of 1.1 mas. This is consistent with zero, given the $\sim$2.5 mas uncertainty of the position measurements. Using this positional uncertainty and the $\sim$4.3-year time difference between the two observations an upper limit of 47 km/s is calculated for the transverse velocity of the source at 17 kpc - the assumed distance for a pulsar wind nebula. This value is significantly low compared to typical Galactic pulsar velocities obtained from VLBI proper motion measurements that fall within the 100--300 km/s range \cite{brisken} and is another piece of evidence against a Galactic origin for the source.

\section{ACKNOWLEDGMENTS}

This research is supported by grants from the U.S. Department of Energy Office of Science, the U.S. National Science Foundation and the Smithsonian Institution, and by NSERC in Canada. We acknowledge the excellent work of the technical support staff at the Fred Lawrence Whipple Observatory and at the collaborating institutions in the construction and operation of the instrument. The VERITAS Collaboration is grateful to Trevor Weekes for his seminal contributions and leadership in the field of VHE $\gamma$-ray astrophysics, and for his interest in the wider applications of IACTs, which made this study possible.


\clearpage

\nocite{*}
\bibliographystyle{aipnum-cp}%

\begin{thebibliography}{17}%
\makeatletter
\providecommand \@ifxundefined [1]{%
 \@ifx{#1\undefined}
}%
\providecommand \@ifnum [1]{%
 \ifnum #1\expandafter \@firstoftwo
 \else \expandafter \@secondoftwo
 \fi
}%
\providecommand \@ifx [1]{%
 \ifx #1\expandafter \@firstoftwo
 \else \expandafter \@secondoftwo
 \fi
}%
\providecommand \natexlab [1]{#1}%
\providecommand \enquote  [1]{``#1''}%
\providecommand \bibnamefont  [1]{#1}%
\providecommand \bibfnamefont [1]{#1}%
\providecommand \citenamefont [1]{#1}%
\providecommand \href@noop [0]{\@secondoftwo}%
\providecommand \href [0]{\begingroup \@sanitize@url \@href}%
\providecommand \@href[1]{\@@startlink{#1}\@@href}%
\providecommand \@@href[1]{\endgroup#1\@@endlink}%
\providecommand \@sanitize@url [0]{\catcode `\$12\catcode `\&12\catcode
  `\#12\catcode `\^12\catcode `\_12\catcode `\%12\relax}%
\providecommand \@@startlink[1]{}%
\providecommand \@@endlink[0]{}%
\providecommand \url  [0]{\begingroup\@sanitize@url \@url }%
\providecommand \@url [1]{\endgroup\@href {#1}{\urlprefix }}%
\providecommand \urlprefix  [0]{URL }%
\providecommand \Eprint [0]{\href }%
\providecommand \doibase [0]{http://dx.doi.org/}%
\providecommand \selectlanguage [0]{\@gobble}%
\providecommand \bibinfo  [0]{\@secondoftwo}%
\providecommand \bibfield  [0]{\@secondoftwo}%
\providecommand \translation [1]{[#1]}%
\providecommand \BibitemOpen [0]{}%
\providecommand \bibitemStop [0]{}%
\providecommand \bibitemNoStop [0]{.\EOS\space}%
\providecommand \EOS [0]{\spacefactor3000\relax}%
\providecommand \BibitemShut  [1]{\csname bibitem#1\endcsname}%
\let\auto@bib@innerbib\@empty
\bibitem [{\citenamefont {{Konigl}}(1981)}]{konigl}%
  \BibitemOpen
  \bibfield  {author} {\bibinfo {author} {\bibfnamefont {A.}~\bibnamefont
  {{Konigl}}},\ }\href {\doibase 10.1086/158638} {\bibfield  {journal}
  {\bibinfo  {journal} {\apj}\ }\textbf {\bibinfo {volume} {243}},\ \unskip\
  \bibinfo {pages} {700--709}February (\bibinfo {year} {1981})}\BibitemShut
  {NoStop}%
\bibitem [{\citenamefont {{Costamante}}\ \emph {et~al.}(2001)\citenamefont
  {{Costamante}}, \citenamefont {{Ghisellini}}, \citenamefont {{Giommi}},
  \citenamefont {{Tagliaferri}}, \citenamefont {{Celotti}}, \citenamefont
  {{Chiaberge}}, \citenamefont {{Fossati}}, \citenamefont {{Maraschi}},
  \citenamefont {{Tavecchio}}, \citenamefont {{Treves}},\ and\ \citenamefont
  {{Wolter}}}]{cost}%
  \BibitemOpen
  \bibfield  {author} {\bibinfo {author} {\bibfnamefont {L.}~\bibnamefont
  {{Costamante}}}, \bibinfo {author} {\bibfnamefont {G.}~\bibnamefont
  {{Ghisellini}}}, \bibinfo {author} {\bibfnamefont {P.}~\bibnamefont
  {{Giommi}}}, \bibinfo {author} {\bibfnamefont {G.}~\bibnamefont
  {{Tagliaferri}}}, \bibinfo {author} {\bibfnamefont {A.}~\bibnamefont
  {{Celotti}}}, \bibinfo {author} {\bibfnamefont {M.}~\bibnamefont
  {{Chiaberge}}}, \bibinfo {author} {\bibfnamefont {G.}~\bibnamefont
  {{Fossati}}}, \bibinfo {author} {\bibfnamefont {L.}~\bibnamefont
  {{Maraschi}}}, \bibinfo {author} {\bibfnamefont {F.}~\bibnamefont
  {{Tavecchio}}}, \bibinfo {author} {\bibfnamefont {A.}~\bibnamefont
  {{Treves}}}, \ and\ \bibinfo {author} {\bibfnamefont {A.}~\bibnamefont
  {{Wolter}}},\ }\href {\doibase 10.1051/0004-6361:20010412} {\bibfield
  {journal} {\bibinfo  {journal} {\aap}\ }\textbf {\bibinfo {volume} {371}},\
  \unskip\ \bibinfo {pages} {512--526}May (\bibinfo {year} {2001})},\ \Eprint
  {http://arxiv.org/abs/astro-ph/0103343} {astro-ph/0103343} \BibitemShut
  {NoStop}%
\bibitem [{\citenamefont {{Piner}}\ and\ \citenamefont {{Edwards}}(2014)}]{PE}%
  \BibitemOpen
  \bibfield  {author} {\bibinfo {author} {\bibfnamefont {B.~G.}\ \bibnamefont
  {{Piner}}}\ and\ \bibinfo {author} {\bibfnamefont {P.~G.}\ \bibnamefont
  {{Edwards}}},\ }\href {\doibase 10.1088/0004-637X/797/1/25} {\bibfield
  {journal} {\bibinfo  {journal} {\apj}\ }\textbf {\bibinfo {volume} {797}},\
  p.~\bibinfo {pages} {25}December (\bibinfo {year} {2014})},\ \Eprint
  {http://arxiv.org/abs/1410.0730} {arXiv:1410.0730 [astro-ph.HE]} \BibitemShut
  {NoStop}%
\bibitem [{\citenamefont {{Meyer}}\ \emph {et~al.}(2011)\citenamefont
  {{Meyer}}, \citenamefont {{Fossati}}, \citenamefont {{Georganopoulos}},\ and\
  \citenamefont {{Lister}}}]{meyer}%
  \BibitemOpen
  \bibfield  {author} {\bibinfo {author} {\bibfnamefont {E.~T.}\ \bibnamefont
  {{Meyer}}}, \bibinfo {author} {\bibfnamefont {G.}~\bibnamefont {{Fossati}}},
  \bibinfo {author} {\bibfnamefont {M.}~\bibnamefont {{Georganopoulos}}}, \
  and\ \bibinfo {author} {\bibfnamefont {M.~L.}\ \bibnamefont {{Lister}}},\
  }\href {\doibase 10.1088/0004-637X/740/2/98} {\bibfield  {journal} {\bibinfo
  {journal} {\apj}\ }\textbf {\bibinfo {volume} {740}},\ p.~\bibinfo {pages}
  {98}October (\bibinfo {year} {2011})},\ \Eprint
  {http://arxiv.org/abs/1107.5105} {arXiv:1107.5105 [astro-ph.CO]} \BibitemShut
  {NoStop}%
\bibitem [{\citenamefont {{Tavecchio}}\ \emph {et~al.}(2011)\citenamefont
  {{Tavecchio}}, \citenamefont {{Ghisellini}}, \citenamefont {{Bonnoli}},\ and\
  \citenamefont {{Foschini}}}]{tevecchio}%
  \BibitemOpen
  \bibfield  {author} {\bibinfo {author} {\bibfnamefont {F.}~\bibnamefont
  {{Tavecchio}}}, \bibinfo {author} {\bibfnamefont {G.}~\bibnamefont
  {{Ghisellini}}}, \bibinfo {author} {\bibfnamefont {G.}~\bibnamefont
  {{Bonnoli}}}, \ and\ \bibinfo {author} {\bibfnamefont {L.}~\bibnamefont
  {{Foschini}}},\ }\href {\doibase 10.1111/j.1365-2966.2011.18657.x} {\bibfield
   {journal} {\bibinfo  {journal} {\mnras}\ }\textbf {\bibinfo {volume}
  {414}},\ \unskip\ \bibinfo {pages} {3566--3576}July (\bibinfo {year}
  {2011})},\ \Eprint {http://arxiv.org/abs/1009.1048} {arXiv:1009.1048
  [astro-ph.HE]} \BibitemShut {NoStop}%
\bibitem [{\citenamefont {{Tanaka}}\ \emph {et~al.}(2014)\citenamefont
  {{Tanaka}}, \citenamefont {{Stawarz}}, \citenamefont {{Finke}}, \citenamefont
  {{Cheung}}, \citenamefont {{Dermer}}, \citenamefont {{Kataoka}},
  \citenamefont {{Bamba}}, \citenamefont {{Dubus}}, \citenamefont {{De
  Naurois}}, \citenamefont {{Wagner}}, \citenamefont {{Fukazawa}},\ and\
  \citenamefont {{Thompson}}}]{tanaka14}%
  \BibitemOpen
  \bibfield  {author} {\bibinfo {author} {\bibfnamefont {Y.~T.}\ \bibnamefont
  {{Tanaka}}}, \bibinfo {author} {\bibfnamefont {{\L}.}~\bibnamefont
  {{Stawarz}}}, \bibinfo {author} {\bibfnamefont {J.}~\bibnamefont {{Finke}}},
  \bibinfo {author} {\bibfnamefont {C.~C.}\ \bibnamefont {{Cheung}}}, \bibinfo
  {author} {\bibfnamefont {C.~D.}\ \bibnamefont {{Dermer}}}, \bibinfo {author}
  {\bibfnamefont {J.}~\bibnamefont {{Kataoka}}}, \bibinfo {author}
  {\bibfnamefont {A.}~\bibnamefont {{Bamba}}}, \bibinfo {author} {\bibfnamefont
  {G.}~\bibnamefont {{Dubus}}}, \bibinfo {author} {\bibfnamefont
  {M.}~\bibnamefont {{De Naurois}}}, \bibinfo {author} {\bibfnamefont {S.~J.}\
  \bibnamefont {{Wagner}}}, \bibinfo {author} {\bibfnamefont {Y.}~\bibnamefont
  {{Fukazawa}}}, \ and\ \bibinfo {author} {\bibfnamefont {D.~J.}\ \bibnamefont
  {{Thompson}}},\ }\href {\doibase 10.1088/0004-637X/787/2/155} {\bibfield
  {journal} {\bibinfo  {journal} {\apj}\ }\textbf {\bibinfo {volume} {787}},\
  p.\ \bibinfo {pages} {155}June (\bibinfo {year} {2014})},\ \Eprint
  {http://arxiv.org/abs/1404.3727} {arXiv:1404.3727 [astro-ph.HE]} \BibitemShut
  {NoStop}%
\bibitem [{\citenamefont {{Arlen}}\ \emph {et~al.}(2014)\citenamefont
  {{Arlen}}, \citenamefont {{Vassilev}}, \citenamefont {{Weisgarber}},
  \citenamefont {{Wakely}},\ and\ \citenamefont {{Yusef Shafi}}}]{arlen}%
  \BibitemOpen
  \bibfield  {author} {\bibinfo {author} {\bibfnamefont {T.~C.}\ \bibnamefont
  {{Arlen}}}, \bibinfo {author} {\bibfnamefont {V.~V.}\ \bibnamefont
  {{Vassilev}}}, \bibinfo {author} {\bibfnamefont {T.}~\bibnamefont
  {{Weisgarber}}}, \bibinfo {author} {\bibfnamefont {S.~P.}\ \bibnamefont
  {{Wakely}}}, \ and\ \bibinfo {author} {\bibfnamefont {S.}~\bibnamefont
  {{Yusef Shafi}}},\ }\href {\doibase 10.1088/0004-637X/796/1/18} {\bibfield
  {journal} {\bibinfo  {journal} {\apj}\ }\textbf {\bibinfo {volume} {796}},\
  p.~\bibinfo {pages} {18}November (\bibinfo {year} {2014})},\ \Eprint
  {http://arxiv.org/abs/1210.2802} {arXiv:1210.2802 [astro-ph.HE]} \BibitemShut
  {NoStop}%
\bibitem [{\citenamefont {{Tavecchio}}\ and\ \citenamefont
  {{Bonnoli}}(2015)}]{TB}%
  \BibitemOpen
  \bibfield  {author} {\bibinfo {author} {\bibfnamefont {F.}~\bibnamefont
  {{Tavecchio}}}\ and\ \bibinfo {author} {\bibfnamefont {G.}~\bibnamefont
  {{Bonnoli}}},\ }\href@noop {} {\bibfield  {journal} {\bibinfo  {journal}
  {ArXiv e-prints}\ December} (\bibinfo {year} {2015})},\ \Eprint
  {http://arxiv.org/abs/1512.05080} {arXiv:1512.05080 [astro-ph.HE]}
  \BibitemShut {NoStop}%
\bibitem [{\citenamefont {{H.E.S.S.~Collaboration}}(2011)}]{hess11}%
  \BibitemOpen
  \bibfield  {author} {\bibinfo {author} {\bibnamefont
  {{H.E.S.S.~Collaboration}}},\ }\href {\doibase 10.1051/0004-6361/201116545}
  {\bibfield  {journal} {\bibinfo  {journal} {\aap}\ }\textbf {\bibinfo
  {volume} {529}},\ p.\ \bibinfo {pages} {A49}May (\bibinfo {year} {2011})},\
  \Eprint {http://arxiv.org/abs/1103.0763} {arXiv:1103.0763 [astro-ph.HE]}
  \BibitemShut {NoStop}%
\bibitem [{\citenamefont {{Gab{\'a}nyi}}\ \emph {et~al.}(2013)\citenamefont
  {{Gab{\'a}nyi}}, \citenamefont {{Dubner}}, \citenamefont {{Giacani}},
  \citenamefont {{Paragi}}, \citenamefont {{Frey}},\ and\ \citenamefont
  {{Pidopryhora}}}]{gabanyi13}%
  \BibitemOpen
  \bibfield  {author} {\bibinfo {author} {\bibfnamefont {K.~{\'E}.}\
  \bibnamefont {{Gab{\'a}nyi}}}, \bibinfo {author} {\bibfnamefont
  {G.}~\bibnamefont {{Dubner}}}, \bibinfo {author} {\bibfnamefont
  {E.}~\bibnamefont {{Giacani}}}, \bibinfo {author} {\bibfnamefont
  {Z.}~\bibnamefont {{Paragi}}}, \bibinfo {author} {\bibfnamefont
  {S.}~\bibnamefont {{Frey}}}, \ and\ \bibinfo {author} {\bibfnamefont
  {Y.}~\bibnamefont {{Pidopryhora}}},\ }\href {\doibase
  10.1088/0004-637X/762/1/63} {\bibfield  {journal} {\bibinfo  {journal}
  {\apj}\ }\textbf {\bibinfo {volume} {762}},\ p.~\bibinfo {pages} {63}January
  (\bibinfo {year} {2013})},\ \Eprint {http://arxiv.org/abs/1110.5039}
  {arXiv:1110.5039} \BibitemShut {NoStop}%
\bibitem [{\citenamefont {{Peter}}\ \emph {et~al.}(2014)\citenamefont
  {{Peter}}, \citenamefont {{Domainko}}, \citenamefont {{Sanchez}},
  \citenamefont {{van der Wel}},\ and\ \citenamefont
  {{G{\"a}ssler}}}]{peter14}%
  \BibitemOpen
  \bibfield  {author} {\bibinfo {author} {\bibfnamefont {D.}~\bibnamefont
  {{Peter}}}, \bibinfo {author} {\bibfnamefont {W.}~\bibnamefont {{Domainko}}},
  \bibinfo {author} {\bibfnamefont {D.~A.}\ \bibnamefont {{Sanchez}}}, \bibinfo
  {author} {\bibfnamefont {A.}~\bibnamefont {{van der Wel}}}, \ and\ \bibinfo
  {author} {\bibfnamefont {W.}~\bibnamefont {{G{\"a}ssler}}},\ }\href {\doibase
  10.1051/0004-6361/201423807} {\bibfield  {journal} {\bibinfo  {journal}
  {\aap}\ }\textbf {\bibinfo {volume} {571}},\ p.\ \bibinfo {pages}
  {A41}November (\bibinfo {year} {2014})},\ \Eprint
  {http://arxiv.org/abs/1408.6976} {arXiv:1408.6976 [astro-ph.HE]} \BibitemShut
  {NoStop}%
\bibitem [{\citenamefont {{Akiyama}}\ \emph {et~al.}(2016)\citenamefont
  {{Akiyama}}, \citenamefont {{Stawarz}}, \citenamefont {{Tanaka}},
  \citenamefont {{Nagai}}, \citenamefont {{Giroletti}},\ and\ \citenamefont
  {{Honma}}}]{akiyama16}%
  \BibitemOpen
  \bibfield  {author} {\bibinfo {author} {\bibfnamefont {K.}~\bibnamefont
  {{Akiyama}}}, \bibinfo {author} {\bibfnamefont {{\L}.}~\bibnamefont
  {{Stawarz}}}, \bibinfo {author} {\bibfnamefont {Y.~T.}\ \bibnamefont
  {{Tanaka}}}, \bibinfo {author} {\bibfnamefont {H.}~\bibnamefont {{Nagai}}},
  \bibinfo {author} {\bibfnamefont {M.}~\bibnamefont {{Giroletti}}}, \ and\
  \bibinfo {author} {\bibfnamefont {M.}~\bibnamefont {{Honma}}},\ }\href@noop
  {} {\bibfield  {journal} {\bibinfo  {journal} {ArXiv e-prints}\ March}
  (\bibinfo {year} {2016})},\ \Eprint {http://arxiv.org/abs/1603.00877}
  {arXiv:1603.00877 [astro-ph.HE]} \BibitemShut {NoStop}%
\bibitem [{\citenamefont {{Straal}}\ \emph {et~al.}(2016)\citenamefont
  {{Straal}}, \citenamefont {{Gabanyi}}, \citenamefont {{van Leeuwen}},
  \citenamefont {{Clarke}}, \citenamefont {{Dubner}}, \citenamefont {{Frey}},
  \citenamefont {{Giacani}},\ and\ \citenamefont {{Paragi}}}]{straal16}%
  \BibitemOpen
  \bibfield  {author} {\bibinfo {author} {\bibfnamefont {S.~M.}\ \bibnamefont
  {{Straal}}}, \bibinfo {author} {\bibfnamefont {K.~E.}\ \bibnamefont
  {{Gabanyi}}}, \bibinfo {author} {\bibfnamefont {J.}~\bibnamefont {{van
  Leeuwen}}}, \bibinfo {author} {\bibfnamefont {T.~E.}\ \bibnamefont
  {{Clarke}}}, \bibinfo {author} {\bibfnamefont {G.}~\bibnamefont {{Dubner}}},
  \bibinfo {author} {\bibfnamefont {S.}~\bibnamefont {{Frey}}}, \bibinfo
  {author} {\bibfnamefont {E.}~\bibnamefont {{Giacani}}}, \ and\ \bibinfo
  {author} {\bibfnamefont {Z.}~\bibnamefont {{Paragi}}},\ }\href@noop {}
  {\bibfield  {journal} {\bibinfo  {journal} {ArXiv e-prints}\ March} (\bibinfo
  {year} {2016})},\ \Eprint {http://arxiv.org/abs/1603.01226}
  {arXiv:1603.01226} \BibitemShut {NoStop}%
\bibitem [{\citenamefont {{Shahinyan}}\ and\ \citenamefont {{VERITAS
  Collaboration}}(2015)}]{shahinyan15}%
  \BibitemOpen
  \bibfield  {author} {\bibinfo {author} {\bibfnamefont {K.}~\bibnamefont
  {{Shahinyan}}}\ and\ \bibinfo {author} {\bibnamefont {{VERITAS
  Collaboration}}},\ }\href@noop {} {\bibfield  {journal} {\bibinfo  {journal}
  {ArXiv e-prints}\ August} (\bibinfo {year} {2015})},\ \Eprint
  {http://arxiv.org/abs/1508.07358} {arXiv:1508.07358 [astro-ph.HE]}
  \BibitemShut {NoStop}%
\bibitem [{\citenamefont {{Holder}}\ and\ \citenamefont {{et
  al.}}(2006)}]{holder06}%
  \BibitemOpen
  \bibfield  {author} {\bibinfo {author} {\bibfnamefont {J.}~\bibnamefont
  {{Holder}}}\ and\ \bibinfo {author} {\bibnamefont {{et al.}}},\ }\href
  {\doibase 10.1016/j.astropartphys.2006.04.002} {\bibfield  {journal}
  {\bibinfo  {journal} {Astroparticle Physics}\ }\textbf {\bibinfo {volume}
  {25}},\ \unskip\ \bibinfo {pages} {391--401}July (\bibinfo {year} {2006})},\
  \Eprint {http://arxiv.org/abs/astro-ph/0604119} {astro-ph/0604119}
  \BibitemShut {NoStop}%
\bibitem [{\citenamefont {{Hovatta}}\ \emph {et~al.}(2014)\citenamefont
  {{Hovatta}}, \citenamefont {{Aller}}, \citenamefont {{Aller}}, \citenamefont
  {{Clausen-Brown}}, \citenamefont {{Homan}}, \citenamefont {{Kovalev}},
  \citenamefont {{Lister}}, \citenamefont {{Pushkarev}},\ and\ \citenamefont
  {{Savolainen}}}]{hovatta14}%
  \BibitemOpen
  \bibfield  {author} {\bibinfo {author} {\bibfnamefont {T.}~\bibnamefont
  {{Hovatta}}}, \bibinfo {author} {\bibfnamefont {M.~F.}\ \bibnamefont
  {{Aller}}}, \bibinfo {author} {\bibfnamefont {H.~D.}\ \bibnamefont
  {{Aller}}}, \bibinfo {author} {\bibfnamefont {E.}~\bibnamefont
  {{Clausen-Brown}}}, \bibinfo {author} {\bibfnamefont {D.~C.}\ \bibnamefont
  {{Homan}}}, \bibinfo {author} {\bibfnamefont {Y.~Y.}\ \bibnamefont
  {{Kovalev}}}, \bibinfo {author} {\bibfnamefont {M.~L.}\ \bibnamefont
  {{Lister}}}, \bibinfo {author} {\bibfnamefont {A.~B.}\ \bibnamefont
  {{Pushkarev}}}, \ and\ \bibinfo {author} {\bibfnamefont {T.}~\bibnamefont
  {{Savolainen}}},\ }\href {\doibase 10.1088/0004-6256/147/6/143} {\bibfield
  {journal} {\bibinfo  {journal} {\aj}\ }\textbf {\bibinfo {volume} {147}},\
  p.\ \bibinfo {pages} {143}June (\bibinfo {year} {2014})},\ \Eprint
  {http://arxiv.org/abs/1404.0014} {arXiv:1404.0014} \BibitemShut {NoStop}%
\bibitem [{\citenamefont {{Brisken}}\ \emph {et~al.}(2003)\citenamefont
  {{Brisken}}, \citenamefont {{Fruchter}}, \citenamefont {{Goss}},
  \citenamefont {{Herrnstein}},\ and\ \citenamefont {{Thorsett}}}]{brisken}%
  \BibitemOpen
  \bibfield  {author} {\bibinfo {author} {\bibfnamefont {W.~F.}\ \bibnamefont
  {{Brisken}}}, \bibinfo {author} {\bibfnamefont {A.~S.}\ \bibnamefont
  {{Fruchter}}}, \bibinfo {author} {\bibfnamefont {W.~M.}\ \bibnamefont
  {{Goss}}}, \bibinfo {author} {\bibfnamefont {R.~M.}\ \bibnamefont
  {{Herrnstein}}}, \ and\ \bibinfo {author} {\bibfnamefont {S.~E.}\
  \bibnamefont {{Thorsett}}},\ }\href {\doibase 10.1086/379559} {\bibfield
  {journal} {\bibinfo  {journal} {\aj}\ }\textbf {\bibinfo {volume} {126}},\
  \unskip\ \bibinfo {pages} {3090--3098}December (\bibinfo {year} {2003})},\
  \Eprint {http://arxiv.org/abs/astro-ph/0309215} {astro-ph/0309215}
  \BibitemShut {NoStop}%
\end{thebibliography}%
\begin{bibliography}{HESSJ1943_gamma16}
\end{bibliography}

\end{document}